\documentclass{article}

\usepackage[final]{neurips_2023}

\usepackage[utf8]{inputenc} 
\usepackage[T1]{fontenc}    
\usepackage{hyperref}       
\usepackage{url}            
\usepackage{booktabs}       
\usepackage{amsfonts}       
\usepackage{nicefrac}       
\usepackage{microtype}      
\usepackage{xcolor}         
\usepackage{comment}
\usepackage{bm}
\usepackage{enumitem}
\usepackage{amsmath}        
\usepackage{mathtools}
\usepackage{wrapfig, lipsum}
\usepackage{adjustbox}

\usepackage{graphicx}
\usepackage{multirow, booktabs}
\usepackage[small, bf]{caption}
\usepackage{subcaption}
\usepackage{makecell}

\DeclarePairedDelimiterX{\infdivx}[2]{\{}{\}}{%
	#1\;\delimsize\|\;#2%
}

\usepackage[
    style=numeric, 
    maxbibnames=99,
    sorting=nyt,
]{biblatex}
\title{Prompting Large Language Models with Speech Recognition Abilities}

\author{%
  Yassir Fathullah$^{1, 2}$\thanks{Work done during internship at Meta AI.} \quad  Chunyang Wu$^{1}$  \quad
  Egor Lakomkin$^{1}$ \quad Junteng Jia$^{1}$  \\ 
  \textbf{Yuan Shangguan}$^{1}$  \quad \textbf{Ke Li}$^{1}$  \quad \textbf{Jinxi Guo}$^{1}$  \quad \textbf{Wenhan Xiong}$^{1}$ \\ \textbf{Jay Mahadeokar}$^{1}$ \quad \textbf{Ozlem Kalinli}$^{1}$  \quad \textbf{Christian Fuegen}$^{1}$  \quad \textbf{Mike Seltzer}$^{1}$
  \\
  Meta AI$^{1}$, University of Cambridge$^{2}$\\
  \texttt{yf286@cam.ac.uk}, \texttt{chunyang@meta.com}
}

\makeatletter
\def\ps@myheadings{%
    \let\@oddfoot\@empty\let\@evenfoot\@empty
    \def\@evenhead{\thepage\hfil\slshape\leftmark}%
    \def\@oddhead{{\slshape\rightmark}\hfil\thepage}%
    \let\@mkboth\@gobbletwo
    \let\sectionmark\@gobble
    \let\subsectionmark\@gobble
    }
  \if@titlepage
  \renewcommand\maketitle{\begin{titlepage}%
  \let\footnotesize\small
  \let\footnoterule\relax
  \let \footnote \thanks
  \null\vfil
  \vskip 60\p@
  \begin{center}%
    {\LARGE \@title \par}%
    \vskip 3em%
    {\large
     \lineskip .75em%
      \begin{tabular}[t]{c}%
        \@author
      \end{tabular}\par}%
      \vskip 1.5em%
    {\large \@date \par}
  \end{center}\par
  \@thanks
  \vfil\null
  \end{titlepage}%
  \setcounter{footnote}{0}%
}
\else
\renewcommand\maketitle{\par
  \begingroup
    \renewcommand\thefootnote{\@fnsymbol\c@footnote}%
    \def\@makefnmark{\rlap{\@textsuperscript{\normalfont\@thefnmark}}}%
    \long\def\@makefntext##1{\parindent 1em\noindent
            \hb@xt@1.8em{%
                \hss\@textsuperscript{\normalfont\@thefnmark}}##1}%
    \if@twocolumn
      \ifnum \col@number=\@ne
        \@maketitle
      \else
        \twocolumn[\@maketitle]%
      \fi
    \else
      \newpage
      \global\@topnum\z@   
      \@maketitle
    \fi
    \thispagestyle{plain}\@thanks
  \endgroup
  \setcounter{footnote}{0}%
}
\makeatother

\begin{document}

\maketitle

\begin{abstract}
    Large language models have proven themselves highly flexible, able to solve a wide range of generative tasks, such as abstractive summarization and open-ended question answering. In this paper we extend the capabilities of LLMs by directly attaching a small audio encoder allowing it to perform speech recognition. By directly prepending a sequence of audial embeddings to the text token embeddings, the LLM can be converted to an automatic speech recognition (ASR) system, and be used in the exact same manner as its textual counterpart. Experiments on Multilingual LibriSpeech (MLS) show that incorporating a conformer encoder into the open sourced LLaMA-7B allows it to outperform monolingual baselines by 18\% and perform multilingual speech recognition despite LLaMA being trained overwhelmingly on English text. Furthermore, we perform ablation studies to investigate whether the LLM can be completely frozen during training to maintain its original capabilities, scaling up the audio encoder, and increasing the audio encoder striding to generate fewer embeddings. The results from these studies show that multilingual ASR is possible even when the LLM is frozen or when strides of almost 1 second are used in the audio encoder opening up the possibility for LLMs to operate on long-form audio.
\end{abstract}

\section{Introduction}
\label{sec:intro}

Large language models (LLMs) \cite{brown2020gpt, chowdhery2022palm, touvron2023llama, scao2022bloom} have been proven to be highly flexible models able to solve a wide range of tasks. By being trained to predict the next token on a vast amount of unsupervised text data, these systems learn to encode world knowledge in the network parameters, useful in many downstream open-domain generative tasks such as abstractive summarization, question answering, knowledge retrieval, text generation and machine translation. 

However, interacting with LLMs purely through text can in many cases be limiting. There exists many other structured modalities which encode information that is difficult to capture through text. For example, audio can encode a wide range of emotions in a person's speech and images can represent the geometry and location of objects that might be much harder to describe through text. Recently published work have extended LLMs with the ability to ingest other modalities. The multi-modal PaLM-E \cite{driess2023palm} combined a large pretrained visual transformer \cite{dehghani2023scaling} with the PaLM LLM \cite{chowdhery2022palm} and were able to achieve state-of-the-art performance on their robotics tasks. Similarly, the work of \cite{zhu2023minigpt} utilize a pretrained visual model and the large language model Vicuna, a derivative of LLaMA \cite{chiang2023vicuna} in creating an aligned model with the ability to reason with both visual and textual inputs. Furthermore \cite{gong2023listen} propose LTU, an extension of LLaMA with an aligned audio encoder trained on an audio question answering corpus, enabling it to reason with and understand sounds. However, LTU has limited speech understanding and recognition abilities.

Due to the immense number of parameters in these large language model oriented systems, it can often be computationally impractical and expensive to adapt the whole system to new tasks. The work of \cite{zhu2023minigpt} trained a single projection layer which adapts the outputs of the visual encoder to be aligned to the language model, representing a highly parameter efficient approach. However, this severely limits the adaptability and performance of the system on new tasks. On the contrary, the multi-modal PaLM-E \cite{driess2023palm} investigated training the whole visual encoder and language model jointly. However, adapting the whole language model is extremely expensive and impractical. Alternative approaches include: inserting adapter layers \cite{rebuffi2017adapter, houlsby2019adapter} or prefix embeddings \cite{li2021prefix} which are trained on the new task. While these approaches are effective parameter efficient approaches they increase the inference costs. Low-rank Adaptation \cite{hu2022lora} solves these issues by using low-rank matrices to modify some parameters of the system and has been shown to be highly promising. The approach is memory efficient during training and does not impact inference runtime.

\textit{Contributions:} In this paper we investigate equipping a large language model with speech recognition abilities by conditioning the LLM on a variable length sequence of audio embeddings. We show that a decoder-only large language model conditioned on the audio sequence is able to perform multilingual speech recognition, outperforming monolingual supervised trained baselines. Furthermore, this paper explores a range of factors that can enable better recognition performance such as the audio encoder model size and frame rate, low-rank adaptation of LLM parameters, text token masking and the type of large language model. Finally, by analysing the outputs of the audio encoder, we show that the audio embeddings are similar and aligned to the text tokens.

\section{Methodology}
\label{sec:methodology}

Our approach will be centered around the use of a large language model (LLM) to model sequences of embeddings irrespective of the modality of the embedding. Inspired by the work of \cite{driess2023palm, zhu2023minigpt} which utilize a visual encoder to generate a \textit{fixed-length} sequence of visual embeddings in the same space as text embeddings, we utilize a pretrained audio encoder to generate a \textit{variable-length} sequence of audial embeddings. By conditioning on the audial embeddings, the large language model can be allowed to perform speech recognition and other speech based tasks. Therefore, the only marginal difference between a traditional LLM and the proposal is the mixing of embeddings of different modalities.

\subsection{Audial Embeddings} 
\label{ssec:audial}

\begin{figure}[h!]
    \centering
    \includegraphics[width=0.9\textwidth]{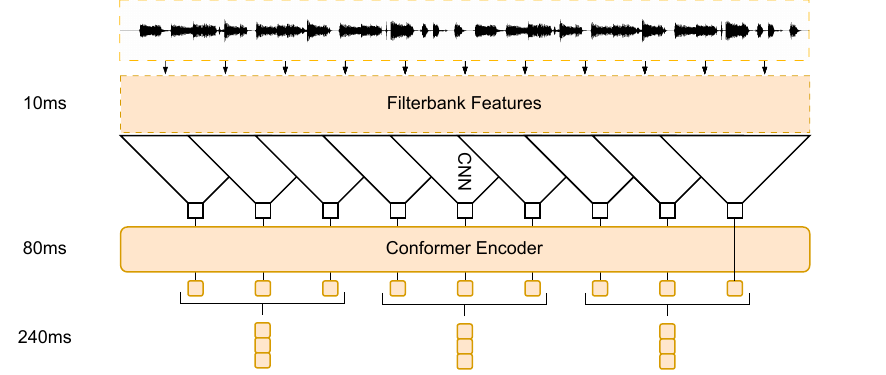}
    \caption{Audio encoder architecture. The initial conformer is trained on a CTC loss. Thereafter the outputs are stacked and projected to the dimension of the LLM to ensure compatibility. This figure showcases a stacking factor of 3 resulting in 240ms embeddings.}
    \label{fig:audio}
\end{figure}

We use a conformer based audio encoder to produce a sequence of embeddings that will be used to condition the LLM similar to a prompt, however, in embeddings space. To ensure the audio encoder can extract useful embeddings it will initially be trained on a simple connectionist temporal classification (CTC) loss. Since the sequence output of this encoder can be very long, one can further reduce the length by stacking consecutive embeddings, resulting in larger but fewer embeddings, see Figure \ref{fig:audio} for the encoder structure.
In this work we investigate different levels of stacking, ranging up to embeddings that encode 960ms of audio which on average contains several tokens worth of information in a single vector. The stacked embeddings are then projected to the hidden dimension of the large language model to ensure they can be prepended to the text embeddings.

\subsection{Large Language Model} 
\label{ssec:textual}

\begin{figure*}[h!]
    \centering
    \includegraphics[width=0.9\textwidth]{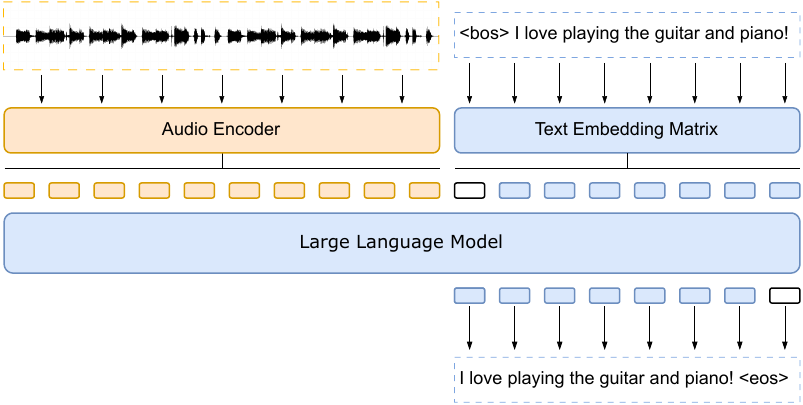}
    \caption{Model architecture. The embedding sequence generated from the audio encoder is directly prepended to the text embeddings sequence. This is directly fed into the decoder-only LLM, tasked with predicting the next token. The LLM can be frozen, adapted with parameter efficient approaches such as LoRA or fully finetuned. This work will investigate the former two.}
    \label{fig:model}
\end{figure*}

Most experiments will utilize the smallest LLaMA-7B model \cite{touvron2023llama}. The causal self-attention parameters of this system will be adapted using a parameter efficient Low-rank Adaptation (LoRA) \cite{hu2022lora}, keeping all other parameters frozen. In an ablation we will investigate whether any LLM parameters need to be tuned at all to perform ASR. Furthermore, we investigate whether the choice of LLM is important by replacing LLaMA with various BLOOM models \cite{scao2022bloom}. The ASR-LLM problem can possibly be reinterpreted as a copying/translation task where the LLM needs to regurgitate the information in the audio sequence. If the audio encoder provides a sequence of embeddings aligned with the text embeddings the problem collapses to a repetition task which should not require the full capacity of an LLM. This interpretation will be investigated in Section \ref{sec:embeddings}. See Figure \ref{fig:model} for an overview of the system.

\section{Experimental Evaluation}
\label{sec:experiments}

\subsection{Dataset}
\label{ssec:dataset}

The Multilingual LibriSpeech (MLS) is a 50k hour ASR corpus derived from read audiobooks of LibriVox \cite{pratap2020mls}. Consisting of 8 languages: English (en), German (de), Dutch (nl), French (fr), Spanish (es), Italian (it), Portuguese (pt) and Polish (pl) the dataset is predominately in English with 44.5k hours. Some low-resource languages such as Portugese and Polish only have 161 and 103 hours respectively. To account for the imbalance in the dataset we follow the strategy outlined in \cite{conneau2021xlsr, babu2021xls} by oversampling from the lower resource languages. Each utterance is up to 20 seconds long. None of our reported word error rates include the use of the n-gram models provided by MLS.

\subsection{Model Setup \& Training Details}
\label{ssec:training}

\textbf{Audio Encoder}\hspace{1pt} The audio encoder operates on 80-d filterbank features with 10ms frame rate. It consists of convolutional feature extractor with a coarse effective stride of 8 followed by linear layer to project the output to 512 dimensions and 18 layers of non-macaron Conformer blocks. The blocks have a hidden dimension of 512, a feed-forward net dimension of 2048, a convolutional kernel size of 11 and 8 attention heads. A final linear layer is used to pretrain the audio encoder using a CTC loss with a SentencePiece \cite{kudo2018sentencepiece} vocabulary of size 1547. The final linear layer is discarded after pretraining. Note that the effectiveness of this relatively small audio encoder of 72 million parameters could be significantly improved by scaling the size up, reducing the level of striding and utilizing a range of unsupervised and semi-supervised learning approaches \cite{conneau2021xlsr, babu2021xls, schneider2019wav2vec, baevski2020wav2vec2, bai2022just, chiu2022rq, chung2021w2vbert}. However, we restrict ourselves to a simpler setup and only use supervised learning to train our models. We focus our attention on showing that an LLM can be conditioned to perform speech recognition and investigate what factors improve its ability at performing this task. 

\textbf{Audial Embeddings}\hspace{1pt} The output of the encoder is a sequence of 512-d vectors with a frame rate of 80ms. To reduce sequence length and memory consumption, every $n$ consecutive frames are stacked to form $512n$-dimensional frames which are projected to 4096-d embeddings to match the LLaMA-7B dimension, with a resulting frame rate of $80n$ms. We investigate producing embeddings up to a frame rate of 960ms, corresponding to stacking 12 consecutive frames. These embeddings are prepended to the text embeddings (as specified in Figure \ref{fig:model}) and fed into the LLM, which is tasked with predicting the next text based token.

\textbf{Large Language Model Adaptation}\hspace{1pt} We use the Low-rank adaptation (LoRA) approach to adapt the key, query, value and output layers of the self-attention mechanism leaving feed-forward nets, embedding and final linear output layer unchanged. Unless specified otherwise, default LoRA hyperparameters are set to a rank of $R = 8$ and $\alpha = 16$. We investigate the impact of $R$ in an ablation study.

\textbf{Training}\hspace{1pt} The audio encoders were initially trained using the Adam optimizer with $\beta_1$ = 0.9, $\beta_2$ = 0.98 \cite{kingma2015adam}. The learning rate was linearly warmed up over 20k training steps up to a peak value of 1e-3 followed by a exponential decaying schedule. This was done on 16 NVIDIA A100 40GBs with 4 gradient accumulations using a per-gpu batch size of up to 500 seconds of audio. The checkpoint with the best validation loss was picked. The joint system with audio encoder and LLM was thereafter trained with a similar schedule of 5k warmup steps up to a peak learning rate of 5e-4 decaying down to 5e-6 over 250k steps. Training was often stopped early withing 100k steps. This was performed on 64 NVIDIA A100 40GBs with 4 gradient accumulations steps using batch sizes of up to 80 seconds. The checkpoint with the lowest validation loss was picked for evaluation.

\textbf{Evaluation}\hspace{1pt} All reported word error rates (WER) exclude the use of external language models provided by \cite{pratap2020mls}. Decoding is done using greedy search with a maximum output token length of 200.

\begin{table*}[b!]
    \centering
    \caption{Language specific and average WER performance on the MLS dataset. The first block \textit{monolingual models} refers to training a separate model for each language. The second block \textit{multilingual model} refers to training a single model on all languages concurrently. The last block refers to pretraining a model on all languages, followed by finetuning a pretrained checkpoint for each language separately.}
    \footnotesize
    \begin{adjustbox}{center}
        \begin{tabular}{l|ccccccccc|c}
            \toprule
            & trainable & \multirow{2}{*}{en} & \multirow{2}{*}{de} & \multirow{2}{*}{nl} & \multirow{2}{*}{fr} & \multirow{2}{*}{es} & \multirow{2}{*}{it} & \multirow{2}{*}{pt} & \multirow{2}{*}{pl} & \multirow{2}{*}{Avg} \\
            & params & & & & & & & & & \\
            \midrule
            \textit{supervised learning: monolingual models} & & & & & & & & & \\
            36L Transformer CTC \cite{pratap2020mls} & 0.3B & 6.8 & 7.1 & 13.1 & 6.6 & 6.7 & 11.8 & 20.5 & 21.7 & 11.8 \\
            36L Transformer CTC \cite{pratap2020mls} w/ LM & 0.3B & \textbf{5.9} & \textbf{6.5} & 12.0 & 5.6 & 6.1 & \textbf{10.5} & 19.5 & 20.4 & 10.8 \\
            \midrule
            \textit{supervised learning: multilingual model} & & & & & & & & & \\
            Decoder-only LLaMA-7B (960ms) & 0.10B & 7.6 & 7.4 & 11.9 & 7.0 & 6.1 & 11.4 & 18.6 & 19.1 & 11.1 \\
            Decoder-only LLaMA-7B (480ms) & 0.09B & 7.3 & 7.4 & 11.9 & 6.7 & 6.1 & 11.5 & 18.3 & 17.0 & 10.8 \\
            Decoder-only LLaMA-7B (240ms) & 0.09B & 7.0 & 7.2 & 11.4 & 6.4 & 6.0 & 11.5 & 17.5 & 16.7 & 10.5 \\
            Decoder-only LLaMA-7B (160ms) & 0.08B & 6.9 & 7.0 & 11.3 & 6.2 & 5.4 & 11.6 & 17.4 & \textbf{14.8} & 10.1 \\
            Decoder-only LLaMA-7B (80ms) & 0.08B & 6.2 & 6.7 & \textbf{11.3} & \textbf{5.5} & \textbf{5.2} & 10.8 & \textbf{16.2} & 15.9 & \textbf{9.7} \\
            \midrule
            \midrule
            \textit{self-supervised learning + monolingual finetuning} & & & & & & & & & \\
            w2v2 XLSR-53 w/ LM & 0.3B & - & 7.0 & 10.8 & 7.6 & 6.3 & 10.4 & 14.7 & 17.2 & 10.6 \\
            \bottomrule
        \end{tabular}
    \end{adjustbox}
    \label{tab:main}
\end{table*}

\subsection{Baselines}

Our approach relies solely on supervised learning and so the most relevant baselines are the monolingual models provided by MLS \cite{pratap2020mls}. Since we follow the same data sampling strategy and setup as in \cite{conneau2021xlsr} we will also include the self-supervised XLSR-53 with monolingual finetuning as a baseline. There are many alternative and powerful audio encoders in literature that achieve highly competitive results on the MLS benchmark, while relevant these systems are often trained using self/semi-supervised approaches with significantly more compute and trainable parameters, representing orthogonal contributions to our aims.

\subsection{Main Results}

Since we keep most parameters in the LLM frozen, and make use of a very small audio encoder, our approach has much fewer trainable parameters compared to baselines, see Table \ref{tab:main}. As expected, the Decoder-only LLaMA with the highest frame rate (80ms) outperforms systems with lower frame rate, also outperforming the monolingual models by 18\% and 10\% on average word error rate. Reducing the frame rate degrades performance, however, even systems with large strides  (480/960ms), reducing the original filterbank sequence by a factor of up to 96, are able to compete with the monolingual baselines. These high striding systems could also be one viable avenue for operating on long-form audio, by compressing the audio sequence length orders of magnitude. 

\subsection{Ablation Studies}

\textbf{Larger Audio Encoders}\hspace{1pt} The level of audio encoder striding has a notable impact on the speech recognition ability of LLaMA. Therefore, we also investigate the number of layers in the audio encoder, scaling it from 72 up to 142 million parameters, see Table \ref{tab:depth}.
\begin{table}[h!]
    \centering
    \caption{Investigating the impact of number of layers of the audio encoder on the MLS dataset.}
    \footnotesize
    \begin{adjustbox}{center}
    \begin{tabular}{l|ccccccccc|c}
        \toprule
        & trainable & \multirow{2}{*}{en} & \multirow{2}{*}{de} & \multirow{2}{*}{nl} & \multirow{2}{*}{fr} & \multirow{2}{*}{es} & \multirow{2}{*}{it} & \multirow{2}{*}{pt} & \multirow{2}{*}{pl} & \multirow{2}{*}{Avg} \\
        & params & & & & & & & & & \\
        \midrule
        18L Conformer (240ms) & 0.09B & 7.0 & 7.2 & 11.4 & 6.4 & 6.0 & 11.5 & 17.5 & 16.7 & 10.5 \\
        24L Conformer (240ms) & 0.11B & 6.6 & 6.6 & 10.8 & 5.9 & 5.4 & 11.5 & 14.5 & 16.8 & 9.8 \\
        36L Conformer (240ms) & 0.16B & 6.1 & 6.3 & 11.0 & 5.5 & 4.9 & 11.1 & 15.9 & 16.7 & 9.7 \\
        \bottomrule
    \end{tabular}
    \end{adjustbox}
    \label{tab:depth}
\end{table}
The largest audio encoder with 36 conformer layers and 240ms striding leads to an average WER of 9.7\% matching the performance of the 18 layer audio encoder with 80ms striding. This shows the importance of the audio encoder in generating higher quality embeddings used in conditioning the LLM. 

\textbf{Low-rank Adaptation}\hspace{1pt} All experiments have fixed the low-rank adaptation parameter to $R = 8$ for adjusting the LLaMA self-attention parameters. We further investigate the impact of the LoRA by adjusting $R \in [0, 8, 16, 32]$; setting $R = 0$ is equivalent to completely freezing LLaMA. All experiments in Table \ref{tab:rank} use 240ms striding.
\begin{table}[h!]
    \centering
    \caption{Investigating the impact of rank $R$. Setting $R = 0$ is equivalent to freezing the LLM.}
    \footnotesize
    \begin{adjustbox}{center}
    \begin{tabular}{l|ccccccccc|c}
        \toprule
        & trainable & \multirow{2}{*}{en} & \multirow{2}{*}{de} & \multirow{2}{*}{nl} & \multirow{2}{*}{fr} & \multirow{2}{*}{es} & \multirow{2}{*}{it} & \multirow{2}{*}{pt} & \multirow{2}{*}{pl} & \multirow{2}{*}{Avg} \\
        & params & & & & & & & & & \\
        \midrule
        Decoder-only LLaMA-7B (240ms) $R = 0$ & 0.08B & 7.5 & 7.4 & 12.0 & 6.8 & 5.9 & 11.8 & 18.2 & 17.4 & 10.9 \\
        Decoder-only LLaMA-7B (240ms) $R = 8$ & 0.09B & 7.0 & 7.2 & 11.4 & 6.4 & 6.0 & 11.5 & 17.5 & 16.7 & 10.5 \\
        Decoder-only LLaMA-7B (240ms) $R = 16$ & 0.10B & 6.3 & 6.8 & 11.4 & 5.7 & 5.5 & 10.8 & 16.3 & 15.0 & 9.7 \\
        Decoder-only LLaMA-7B (240ms) $R = 32$ & 0.11B & 6.0 & 6.5 & 11.1 & 5.4 & 5.2 & 10.9 & 15.7 & 15.3 & 9.5 \\
        \bottomrule
    \end{tabular}
    \end{adjustbox}
    \label{tab:rank}
\end{table}
Each rank adds approximately 1 million trainable parameters. Interestingly, keeping LLaMA frozen and only training the audio encoder leads to reasonable results with an average WER of 10.9\%. This would also maintain the original capabilities of the LLM; all other finetuning setups would negatively affect the ability of LLaMA in performing text based tasks \cite{driess2023palm}. Furthermore, increasing the rank of the trainable parameters significantly improves performance, where $R = 32$ is able to achieve an average WER of 9.5\%, outperforming the best system in Table \ref{tab:main} which uses 80ms striding and $R = 8$. Based on these results, parameter tuning the whole LLM could lead to additional performance gains but is significantly more expensive to train.

\textbf{Masking}\hspace{1pt} Since the training task is based on causal next token prediction, but is conditioned on the audio sequence which contains the needed information, masking text tokens could be useful in boosting performance \cite{li2023deliberation}. The table below shows performance when a fraction $F \in [0.000, 0.125, 0.250, 0.375, 0.500]$ of the text tokens are randomly replaced with the \texttt{<unk>} token during training.
\begin{table}[h!]
    \centering
    \caption{Masking a fraction $F$ of text tokens during training.}
    \footnotesize
    \begin{adjustbox}{center}
    \begin{tabular}{l|ccccccccc|c}
        \toprule
        & trainable & \multirow{2}{*}{en} & \multirow{2}{*}{de} & \multirow{2}{*}{nl} & \multirow{2}{*}{fr} & \multirow{2}{*}{es} & \multirow{2}{*}{it} & \multirow{2}{*}{pt} & \multirow{2}{*}{pl} & \multirow{2}{*}{Avg} \\
        & params & & & & & & & & & \\
        \midrule
        Decoder-only LLaMA-7B (240ms) $F = 0.000$ & 0.09B & 7.0 & 7.2 & 11.4 & 6.4 & 6.0 & 11.5 & 17.5 & 16.7 & 10.5 \\
        Decoder-only LLaMA-7B (240ms) $F = 0.125$ & 0.09B & 6.7 & 7.0 & 11.3 & 6.1 & 5.6 & 11.3 & 16.8 & 16.3 & 10.1 \\
        Decoder-only LLaMA-7B (240ms) $F = 0.250$ & 0.09B & 6.5 & 6.9 & 11.3 & 6.1 & 5.6 & 11.2 & 16.5 & 15.1 & 9.9 \\
        Decoder-only LLaMA-7B (240ms) $F = 0.375$ & 0.09B & 6.5 & 7.0 & 11.4 & 6.1 & 5.4 & 11.3 & 17.4 & 16.2 & 10.2 \\
        Decoder-only LLaMA-7B (240ms) $F = 0.500$ & 0.09B & 6.4 & 7.0 & 11.5 & 6.2 & 5.1 & 11.1 & 17.1 & 16.8 & 10.2 \\
        \bottomrule
    \end{tabular}
    \end{adjustbox}
    \label{tab:mask}
\end{table}
The introduction of masked text tokens during training can lead to notable improvements in performance, with $F = 0.250$ leading to a 5.7\% average WER improvement compared to the baseline $F = 0.000$. However, beyond this point, increasing the level of masking has a negative impact on the low resource languages Portuguese and Polish. It is possible to set different levels of masking depending on the amount of language specific data but we leave this investigation to future work.

\textbf{Large Language Model}\hspace{1pt} LLaMA was trained on predominantly English text with a small fraction covering other languages \cite{touvron2023llama}. BLOOM \cite{scao2022bloom}, on the other hand, was specifically designed to be multilingual and has support for an order of magnitude more languages. Therefore, we replace LLaMA-7B with a choice of \{BLOOM-560M, BLOOM-1B7, BLOOM-7B1\} to understand the impact of LLM and how performance changes with increasing LLM scale, see Table \ref{tab:bloom}.
\begin{table}[h!]
    \centering
    \caption{Replacing LLaMA-7B with various BLOOM language models.}
    \footnotesize
    \begin{adjustbox}{center}
    \begin{tabular}{l|ccccccccc|c}
        \toprule
        & trainable & \multirow{2}{*}{en} & \multirow{2}{*}{de} & \multirow{2}{*}{nl} & \multirow{2}{*}{fr} & \multirow{2}{*}{es} & \multirow{2}{*}{it} & \multirow{2}{*}{pt} & \multirow{2}{*}{pl} & \multirow{2}{*}{Avg} \\
        & params & & & & & & & & & \\
        \midrule
        Decoder-only LLaMA-7B (240ms) & 0.09B & 7.0 & 7.2 & 11.4 & 6.4 & 6.0 & 11.5 & 17.5 & 16.7 & 10.5 \\
        \midrule
        Decoder-only BLOOM-560M (240ms) & 0.07B & 8.2 & 8.4 & 12.6 & 7.3 & 6.5 & 12.5 & 18.3 & 19.8 & 11.7 \\
        Decoder-only BLOOM-1B7 (240ms) & 0.08B & 7.5 & 8.3 & 12.2 & 6.7 & 5.8 & 12.2 & 16.6 & 19.0 & 11.0 \\
        Decoder-only BLOOM-7B1 (240ms) & 0.08B & 7.0 & 7.8 & 12.1 & 5.9 & 5.3 & 11.8 & 15.6 & 17.7 & 10.4 \\
        \bottomrule
    \end{tabular}
    \end{adjustbox}
    \label{tab:bloom}
\end{table}
Comparing LLaMA-7B and the similarly sized BLOOM-7B1 we observe no significant difference in average WER. Although BLOOM is multilingual it seems this ability is not as impactful once the system is trained on a multilingual speech dataset. However, there is a clear trend showing significantly better performance from scaling an LLM while keeping the conformer audio encoder fixed.

\section{Analysing Audio Encoder Text Alignment}
\label{sec:embeddings}

As hypothesized in Section \ref{ssec:textual} the speech recognition task can be interpreted as a regurgitation task---the language model is tasked with cleaning and repeating (in the same order) information that is present in the audio encoder output sequence. Since the audio encoder is trained to generate embeddings in the same semantic space as the text embeddings, this implies that the audio and text embeddings should be monotonically aligned for a properly trained system. 

\begin{figure*}[h!]
    \centering
    \begin{subfigure}[b]{0.953\textwidth}
         \centering
         \includegraphics[width=\textwidth]{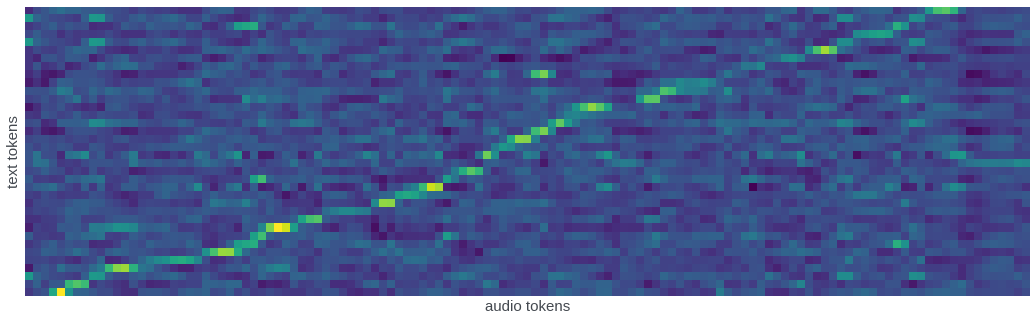}
         \caption{}
         \label{fig:80ms}
     \end{subfigure}
     \begin{subfigure}[b]{0.414\textwidth}
         \centering
         \includegraphics[width=\textwidth]{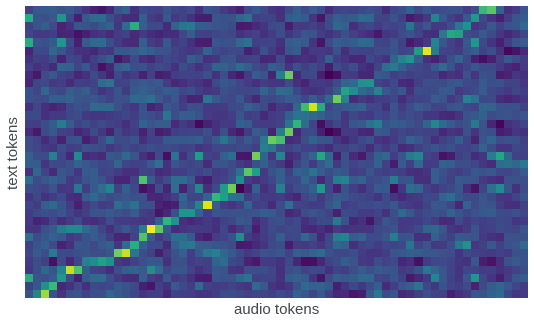}
         \caption{}
         \label{fig:160ms}
     \end{subfigure}
     \begin{subfigure}[b]{0.282785\textwidth}
         \centering
         \includegraphics[width=\textwidth]{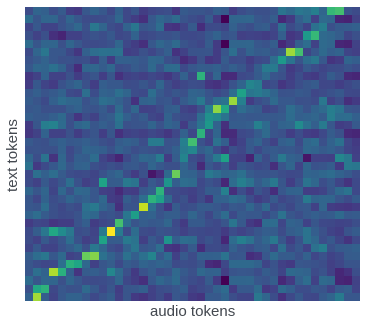}
         \caption{}
         \label{fig:240ms}
     \end{subfigure}
     \begin{subfigure}[b]{0.1495\textwidth}
         \centering
         \includegraphics[width=\textwidth]{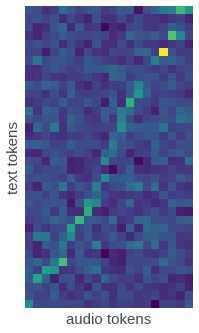}
         \caption{}
         \label{fig:480ms}
     \end{subfigure}
     \begin{subfigure}[b]{0.087285\textwidth}
         \centering
         \includegraphics[width=\textwidth]{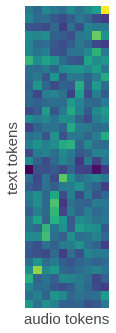}
         \caption{}
         \label{fig:960ms}
     \end{subfigure}
    \caption{The pairwise cosine similarity between every pair of audio and text embeddings for a given test example from the English set. The subfigures \textbf{(a)}-\textbf{(e)} represent the models in Table \ref{tab:main} with stridings ranging from 80ms up to 960ms.}
    \label{fig:alignment}
\end{figure*}

We therefore, compute the cosine similarity between each possible pair of audio and text embedding for an English test set example. This is done for the LLaMA models in \ref{tab:main} to understand the impact of increased striding on the impact of alignment, see Figure \ref{fig:alignment}. 
These alignment plots support the hypothesis that the encoder is attempting to align the audio embeddings to the text in a monotonic manner. As the striding is increase, the task of aligning audio to text becomes harder and harder. Furthermore, this begs the question whether or not the audio encoder can benefit from further supervision by training the output to be monotonically aligned to the text, instead of indirectly training it through next token prediction via the language model.

\section{Conclusion}
\label{sec:conclusion}

Overall this work has shown a simple procedure for enabling multilingual speech recognition with a large language model. By prepending an audio embedding sequence, the large language model can be triggered to perform speech recognition in a decoder-only fashion. Furthermore, this work investigates a range of different factors that are key in enabling better recognition performance including analysing the audio encoder stride \& size. The paper also investigates the importance of the LLM by comparing LLaMA against BLOOM, the importance of tuning the LLM with the use of low-rank adapters and finally how the LLM can perform better recognition by augmenting the input with masking. After joint training of the encoder and LLM it was shown that the audio embeddings are tending to be aligned with the text embeddings. Future work can make use of this observation by directly training the audio encoder to be aligned with the language model.

\clearpage
\AtNextBibliography{\small}
\printbibliography

\end{document}